\documentclass[cameraready]{Interspeech}

\title{StuPASE: Towards Low-Hallucination Studio-Quality \\Generative Speech Enhancement}

\author[affiliation={1,2}, orcid=0009-0006-4373-8306]{Xiaobin}{Rong}
\author[affiliation={1,3}]{Jun}{Gao}
\author[affiliation={1,3}]{Zheng}{Wang}
\author[affiliation={2}]{Mansur}{Yesilbursa}
\author[affiliation={2}]{Kamil}{Wojcicki}
\author[affiliation={1,3}, correspondingauthor]{Jing}{Lu}

\address{
    $^1$ Key Laboratory of Modern Acoustics, Nanjing University \\
    $^2$ Collaboration AI, Cisco Systems, Inc. \\
    $^3$ NJU-Horizon Intelligent Audio Lab, Horizon Robotics
}

\email{xiaobin.rong@smail.nju.edu.cn, lujing@nju.edu.cn}

\usepackage{comment}
\usepackage[table]{xcolor}
\usepackage{multirow}
\usepackage{threeparttable}

\begin{document}

\maketitle

\begin{abstract}
    Achieving high perceptual quality without hallucination remains a challenge in generative speech enhancement (SE). A representative approach, PASE, is robust to hallucination but has limited perceptual quality under adverse conditions. We propose StuPASE, built upon PASE to achieve studio-level quality while retaining its low-hallucination property. First, we show that finetuning PASE with dry targets rather than targets containing simulated early reflections substantially improves dereverberation. Second, to address performance limitations under strong additive noise, we replace the GAN-based generative module in PASE with a flow-matching module, enabling studio-quality generation even under highly challenging conditions. Experiments demonstrate that StuPASE consistently produces perceptually high-quality speech while maintaining low hallucination, outperforming state-of-the-art SE methods. Audio demos are available at: \urlstyle{same}\url{https://xiaobin-rong.github.io/stupase_demo/}.
\end{abstract}

\section{Introduction}
Generative speech enhancement (SE) offers a compelling alternative to discriminative methods by synthesizing speech with superior perceptual quality~\cite{StoRM, FlowSE}. Representative approaches include generative adversarial networks (GANs)~\cite{SEGAN}, diffusion~\cite{StoRM, CDiffSE}, flow-matching~\cite{FlowSE}, and language models (LMs)~\cite{GenSE, LLaSE-G1}. Despite their strengths, generative models are prone to \textit{hallucination}, causing inconsistencies in linguistic content or speaker characteristics relative to the original speech~\cite{LPS, URGENT2025}. 

Recent studies have explored leveraging semantic representations\footnote{Also referred to as \emph{phonetic} representations, as they exhibit properties closer to phoneme-like units~\cite{Phonetic_analysis, SSL_phonetic}.} to improve content accuracy in generative SE~\cite{GenSE, AnyEnhance, SISE, PASE, SenSE}. These approaches typically adopt a two-stage semantic-acoustic paradigm: semantic enhancement first produces purified semantic conditions, which then guide acoustic enhancement to generate clean acoustic representations such as Mel spectrograms~\cite{SenSE} and codec tokens~\cite{GenSE, AnyEnhance, SISE}, or raw waveforms~\cite{PASE}. These two stages target complementary objectives in SE: linguistic integrity and perceptual quality, respectively.

Within this two-stage paradigm, semantic enhancement is realized through diverse strategies, including LMs~\cite{GenSE, SenSE}, MLP-based alignment~\cite{AnyEnhance}, diffusion models~\cite{SISE}, and denoising representation distillation in Phonologically Anchored Speech Enhancer (PASE)~\cite{PASE}. Notably, PASE reduces hallucination efficiently within the self-supervised learning (SSL) speech encoder, without requiring an additional modeling network. 
Regarding acoustic enhancement, PASE employs a simple GAN-based module that allows efficient inference. However, compared with other methods—such as LM-based generative models~\cite{GenSE}, masked generative models~\cite{AnyEnhance}, diffusion models~\cite{SISE}, or flow-matching models~\cite{SenSE}—PASE’s acoustic enhancement produces only moderate-quality speech under challenging conditions, especially with strong reverberation.

To overcome this limitation of PASE, we revisit its training targets, which retain the first 50~ms of reflections—a common practice in discriminative SE~\cite{DRB_TCSA, DRB_RTS, URGENT2024, URGENT2025_rank1_tencent}. However, our analysis reveals that targets with artificially added early reflections could be perceptually reverberant and exhibit blurred spectral details.\footnote{Audio examples are provided on the demo page.} Unlike discriminative SE models that focus on direct signal mapping, generative models aim to learn the full distribution of the target speech. Consequently, distortions in the training targets can bias the learned distribution and affect the quality of generated speech. As our curated speech data already contains minor, naturally occurring reflections sufficient for perception, we adopt \emph{dry} recordings (i.e., without additional simulated reflections) as training targets and demonstrate that finetuning PASE on such targets yields substantial improvements under severe reverberation.

Despite this progress, the finetuned PASE still struggles to generate \emph{studio-quality} speech, which in this work we define as speech free of background noise and reverberation, and sounding natural. Under challenging conditions, PASE outputs fall short of this standard and often exhibit residual noise, reverberation, and processing-induced artifacts.
We attribute these limitations to the inherent constraints of the GAN‑based acoustic enhancement module,~i.e.,~its heavy reliance on noisy conditioning can leave residual noise and reverberation, and its limited generative capacity may cause over‑suppression and artifacts. To address this, we replace it with a more expressive flow‑matching module, enabling studio‑quality speech generation even under highly adverse conditions. Our contributions are threefold:
\begin{itemize}
    \item We reveal the benefit of dry training targets for dereverberation in PASE, as finetuning pre-trained PASE with such targets substantially improves dereverberation.
    \item We introduce \textbf{StuPASE}, which replaces PASE’s GAN-based generative module with a flow-matching module, enabling studio-quality speech generation under adverse conditions.
    \item We show that StuPASE outperforms existing state-of-the-art (SOTA) SE methods, simultaneously achieving low hallucination and studio-level perceptual quality.
\end{itemize}

\section{Method}
\subsection{PASE Overview}
The PASE framework~\cite{PASE} comprises two main components: (i) a semantic enhancement module, denoising WavLM (DeWavLM); and (ii) an acoustic enhancement module, the GAN-based dual-stream vocoder (DualVocoder). DeWavLM is obtained by finetuning a pre-trained WavLM using denoising representation distillation (DRD), in which a student learns to map noisy waveforms to clean \emph{phonetic representations} produced by a frozen teacher from the final transformer layer. This results in enhanced phonetic representations that provide a clean, high-quality encoding of semantic information.
The DualVocoder reconstructs enhanced waveforms from dual-stream representations: (i) the enhanced phonetic representations, and (ii) the low-level acoustic representations from the first transformer layer, which retain fine-grained acoustic details but may still carry noise and reverberation. This dual-stream design aims to synthesize speech that faithfully preserves both the linguistic content and speaker characteristics.

\subsection{Dry-Target Finetuning}
As noted earlier, simulated early reflections may produce perceptually reverberant training targets which could degrade both modules in PASE. For DeWavLM, they can contaminate the phonetic targets from the teacher, biasing the student model and compromising semantic fidelity. For the DualVocoder, they may mislead the model into learning an undesired distribution, thereby impairing the quality of the generated speech.

To overcome these issues, we finetune PASE using dry targets, and refer to the finetuned model as \textbf{PASE-R}. The finetuning is carried out in two stages, corresponding to PASE’s two components. First, we finetune DeWavLM using the same DRD strategy. The teacher and student are initialized from pre-trained WavLM and DeWavLM weights, respectively. The teacher generates target representations from dry speech, and the student is trained to minimize the mean-squared error (MSE) between its output given noisy speech and the teacher’s output. This finetuned model is referred to as \textbf{DeWavLM-R}. Next, we finetune the DualVocoder on top of DeWavLM-R to reconstruct the dry waveform, yielding DualVocoder-R. This encourages the generative module to learn the distribution of studio-quality speech. DeWavLM-R and DualVocoder-R together constitute PASE-R.

\subsection{Flow-Matching-Based Generation}
Another limitation of the PASE framework may stem from its simple GAN-based acoustic enhancement design, which can become a bottleneck in fully suppressing noise and reverberation without introducing artifacts.
To further advance toward studio-level quality, we replace the GAN-based DualVocoder with a more expressive flow-matching module paired with a Mel vocoder, forming StuPASE. 
Unlike PASE's original DualVocoder that directly reconstructs enhanced waveforms from WavLM representations, StuPASE decouples generation by first producing a high-fidelity enhanced Mel spectrogram via flow-matching, which is then converted to a waveform using a Mel vocoder. We adopt flow matching for its high-fidelity generation and lower sampling requirements compared to other diffusion-based approaches~\cite{Solve_inv, Flow}. Consistent with this choice, we use a Mel vocoder, as Mel spectrograms provide a stable domain for flow-matching models in speech synthesis~\cite{E2-TTS, F5-TTS} and enhancement~\cite{FlowSE, SenSE}, and Mel-based vocoders are widely used for high-quality audio synthesis~\cite{HiFiGAN, BigVGAN, Vocos}.

\begin{figure}[t]
    \centering
    \includegraphics[width=\linewidth]{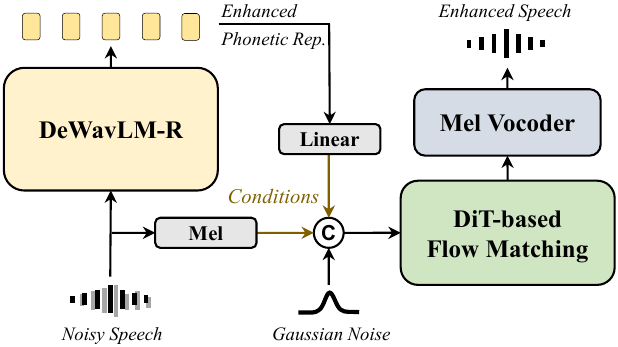}
    \caption{Overview of the proposed StuPASE framework.}
    \label{fig:stupase}
\end{figure}

As illustrated in Fig.~\ref{fig:stupase}, StuPASE comprises three main components: DeWavLM-R, a DiT-based~\cite{DiT} flow-matching module, and a Mel vocoder. Specifically, DeWavLM-R performs semantic enhancement, producing purified phonetic representations that capture semantic information from a noisy waveform. These representations are linearly projected and concatenated with the noisy Mel spectrogram to form conditioning signals, providing both high-level semantic guidance and low-level acoustic structure. The resulting conditions are then concatenated with the Gaussian noise and fed to the flow-matching module, which performs acoustic enhancement, generating a clean Mel spectrogram that matches the distribution of studio-quality speech. Finally, a pre-trained Mel vocoder reconstructs the enhanced waveform.

During training, an additional clean Mel spectrogram is provided as a condition to guide the flow-matching module toward the distribution of clean speech. We further adopt the speech-infilling training paradigm following SenSE~\cite{SenSE}. 
This involves generating a missing clean Mel segment spanning consecutive temporal frames based on: (i) the surrounding context, (ii) the incomplete noisy Mel spectrogram (masked using the same strategy, with regions sampled independently), and (iii) the complete phonetic representation sequence.
The training objective is to minimize the MSE between the predicted and the target velocities within the masked region. This paradigm encourages the flow-matching module to learn the distribution of clean speech by fully exploiting the purified phonetic representations while reducing reliance on low-level noisy acoustic cues, leading to cleaner and more linguistically faithful enhancement.

We highlight that while StuPASE adopts a flow-matching module similar to SenSE, it differs fundamentally in semantic enhancement. Instead of relying on predicted discrete tokens from a semantic speech encoder followed by a large LM, StuPASE directly uses the continuous phonetic representations from DeWavLM-R as semantic conditioning, eliminating the need for an additional semantic modeling network. Our results in Sec.~\ref{sec:comparison} further demonstrate that despite this simplification, StuPASE achieves superior hallucination reduction over SenSE.

\section{Experiments}
\subsection{Datasets}
We construct the training dataset following PASE~\cite{PASE}, using clean speech from LibriVox (DNS5)~\cite{DNS5}, LibriTTS~\cite{LibriTTS}, VCTK~\cite{VCTK}, and Common Voice 19.0~\cite{CommonVoice}, totalling around 2,000 hours. Noise data comes from DNS5, WHAM!~\cite{WHAM}, FSD50K~\cite{FSD50K}, and FMA~\cite{FMA}, while room impulse responses (RIRs) are taken from openSLR26 and openSLR28~\cite{openSLR}. For finetuning PASE, no additional data filtering is applied. For training the flow-matching module and Mel vocoder, we curate studio-quality targets: Common Voice 19.0 is excluded due to its in-the-wild nature, and LibriSpeech~\cite{LibriSpeech} is added with its utterances denoised using a pre-trained DPCRN~\cite{DPCRN}. All speech is filtered with a UTMOS threshold of 4.0, yielding around 1,000 hours of data. All audio is resampled to 16~kHz.

Training mixtures are generated on-the-fly by convolving clean speech with a random RIR (80\% probability) and adding noise at SNRs uniformly sampled from -5 to 15 dB. The training target is dry clean speech without additional early reflections.

For evaluation, we use the DNS1 test set~\cite{DNS1} (\emph{with-reverb} and \emph{no-reverb}, 150 utterances each), using no-reverb clean utterances as references when computing metrics. 
Additionally, we construct a test set of 1,000 noisy–clean paired samples by combining 1,000 speech utterances from the LibriSpeech \emph{test-clean} split with unseen noise clips and RIRs, including 750 from openSLR and 250 simulated high-RT60 RIRs (0.6–1.6 s).

\subsection{Baselines}
We compare StuPASE with SOTA SE models, including the discriminative TF-GridNet~\cite{TF-GridNet}, the generative FlowSE~\cite{FlowSE}, PASE~\cite{PASE}, SenSE~\cite{SenSE}, and the commercial Adobe Enhance Speech V2 (AES-V2)~\cite{AdobePodcastEnhance}. For TF-GridNet, we use the official baseline checkpoint from the URGENT 2025 Challenge~\cite{URGENT2025}. 
For FlowSE, since the provided checkpoint is trained on a small dataset, we retrain it from scratch using the official implementation. PASE and SenSE are evaluated using their official checkpoints. 
AES-V2 inference is performed via the official API.

\subsection{Evaluation Metrics}
We adopt objective metrics covering three key dimensions:
\begin{itemize}
    \item \textbf{Perceptual Quality}: DNSMOS~\cite{DNSMOS-P835} and UTMOS~\cite{UTMOS}.
    \item \textbf{Speaker Fidelity}: Speaker similarity (SpkSim), computed via a finetuned-WavLM-Large-based ECAPA-TDNN~\cite{UniSpeechSpeakerVerification}.
    \item \textbf{Linguistic Integrity}: Levenshtein phoneme similarity (LPS)~\cite{LPS}, SpeechBERTScore (SBS)~\cite{SpeechBERTScore}, and word error rate (WER) with Whisper-Large-v3~\cite{Whisper-Large-v3}. When reference transcripts are unavailable, ASR transcriptions of clean speech are used as pseudo-references, yielding dWER.
\end{itemize}

We further conduct subjective evaluations on 70 samples from the simulated test set (noisy utterances with UTMOS $\le$ 1.3 and SNR $\le$ 5 dB). We report quality mean opinion score (Q-MOS) obtained via standard absolute category rating (ACR) test~\cite{itu_p800_1996} and similarity mean opinion score (S-MOS), following~\cite{Genhancer}. For Q-MOS, listeners rate the quality of the enhanced speech without the clean reference on a 5-point scale (1: Bad, 5: Excellent). For S-MOS, listeners judge the speaker similarity of the processed speech relative to the clean reference using a 5-point scale (1: Completely Different, 5: Identical). All listening tests were conducted on Prolific~\cite{Prolific} with native English speakers, involving 419 listeners for Q-MOS (6 responses/file) and 349 for S-MOS (5 responses/file).

\subsection{Implementation Details}
\begin{itemize}
    \item \textbf{PASE Finetuning}: DeWavLM-R and DualVocoder-R are initialized with their pre-trained weights from PASE, and finetuned for 50k steps with peak learning rate 2e-5. Per-GPU batch sizes are 20 (each segment 4 s) for DeWavLM-R and 24 (each segment 2 s) for DualVocoder-R.

    \item \textbf{Mel Vocoder}: The Mel transformation produces 100-dim log Mel features (window/FFT 1280, hop 320, 50 Hz frame rate). The Mel vocoder follows DualVocoder in PASE, employing the same architecture (i.e., improved Vocos~\cite{Vocos, WavTokenizer}), discriminators, and loss functions. It is trained for 200k steps with peak learning rate 2e-4 and per-GPU batch size 60 (1 s).
    
    \item \textbf{Flow-Matching}: The DiT backbone consists of 12 layers, 16 attention heads, a hidden dimension of 1024, and a feed-forward dimension of 2048. The semantic projection layer maps 1024-dim inputs to 512-dim outputs. Masking ratios are uniformly sampled from $[0.7, 1.0]$ for the clean Mel and $[0.5, 1.0]$ for the noisy Mel. It is trained for 100k steps, with peak learning rate 1e-4 and per-GPU batch size 60 (4 s). Inference uses 8 sampling steps, following SenSE.
\end{itemize}
All models are trained on two NVIDIA 4090 GPUs using AdamW~\cite{AdamW} optimizer, with linear warm-up over the first 10\% of steps followed by cosine decay to 1e-6. 

\begin{table}[t]
\caption{Ablation on DNS1 with-reverb test set. Blue-shaded rows are for references; $^\dagger$denotes inference on clean speech.}
\centering
\setlength{\tabcolsep}{1.5mm}
\resizebox{\linewidth}{!}{
\begin{tabular}{ccccc}
\toprule
Model & UTMOS $\uparrow$ & SpkSim $\uparrow$ & dWER (\%) $\downarrow$ \\ \midrule
\rowcolor{blue!8} 
Noisy & 1.30 & 0.70 & 10.23 \\
\rowcolor{blue!8}
Clean & 4.14 & 1.00 & 0.00 \\ \midrule
DeWavLM & 2.42 & 0.54 & 10.30 \\
DeWavLM-R & 3.98 & 0.55 & 8.69 \\
PASE & 1.61 & 0.74 & 9.78 \\
PASE-R & 3.23 & \textbf{0.80} & 8.01 \\ \midrule
\rowcolor{blue!8}
Mel vocoder$^\dagger$ & 3.85 & 0.96 & 0.92 \\
StuPASE & \textbf{4.01} & 0.74 & \textbf{7.89} \\
StuPASE w/ noisy semantic &3.75 & 0.66 & 19.79 \\
StuPASE w/o semantic & 3.25 & 0.57 & 36.36 \\
StuPASE w/o masking & 3.82 & 0.74 & 8.79 \\ \bottomrule
\end{tabular}
}
\label{tab:ablation}
\end{table}

\begin{table*}[t]
\centering

\caption{Comparison results on the DNS1 test set. \textbf{Bold} indicates the best scores, and \underline{underline} indicates the second-best.}
\setlength{\tabcolsep}{1.5mm}
\resizebox{\linewidth}{!}{
\begin{tabular}{@{}cccccccccccccc@{}}
\toprule
\multirow{2}{*}{Model} & \multicolumn{6}{c}{No-Reverb} & & \multicolumn{6}{c}{With-Reverb} \\ \cmidrule(lr){2-7} \cmidrule(l){9-14} 
 & DNSMOS $\uparrow$ & UTMOS $\uparrow$ & SBS $\uparrow$ & LPS $\uparrow$ & SpkSim $\uparrow$ & dWER (\%) $\downarrow$ &  & DNSMOS $\uparrow$ & UTMOS $\uparrow$ & SBS $\uparrow$ & LPS $\uparrow$& SpkSim $\uparrow$ & dWER (\%) $\downarrow$\\ \midrule
Noisy & 2.48 & 2.36 & 0.80 & 0.90 & 0.96 & 3.51 &  & 1.39 & 1.30 & 0.61 & 0.63 & 0.79 & 10.23 \\
Clean & 3.28 & 4.14 & 1.00 & 1.00 & 1.00 & 0.00 &  & 3.28 & 4.14 & 1.00 & 1.00 & 1.00 & 0.00 \\ \midrule
TF-GridNet \cite{TF-GridNet} & 3.34 & 3.86 & 0.91 & \underline{0.97} & \textbf{0.94} & 2.86 & & 2.63 & 1.42 & 0.77 & 0.88 & \underline{0.80} & \underline{8.86} \\
FlowSE \cite{FlowSE} & \underline{3.38} & 3.76 & 0.90 & 0.94 & 0.89 & 4.65 & & 3.34 & 3.51 & 0.81 & 0.85 & 0.72 & 15.58 \\
PASE \cite{PASE} & 3.29 & 3.95 & \textbf{0.93} & \underline{0.97} & 0.88$^{*}$ & \textbf{2.71} & & 2.75 & 1.61 & 0.81 & \underline{0.90} & 0.74$^{*}$ & 9.78 \\
SenSE \cite{SenSE} & \underline{3.38} & 3.85 & \underline{0.92} & \textbf{0.98} & \underline{0.92} & 5.49 & & 3.37 & 3.55 & \underline{0.85} & \textbf{0.92} & \textbf{0.82} & 11.30 \\
AES-V2 & \textbf{3.42} & \textbf{4.08} & 0.88 & 0.94 & 0.83 & 4.55 & & \textbf{3.40} & \underline{3.71} & 0.70 & 0.79 & 0.63 & 18.87 \\ \midrule
StuPASE &\textbf{ 3.42} & \underline{3.99} & 0.91 & \textbf{0.98} & 0.88 & \underline{2.84} & & \underline{3.39} & \textbf{4.01} &\textbf{ 0.86} & \textbf{0.92} & 0.74 & \textbf{7.89}\\ \bottomrule
\end{tabular}
}

\vspace{1mm}
\begin{minipage}{\linewidth}
\raggedright
\footnotesize
$^{*}$ The SpkSim scores differ from those reported in the original paper \cite{PASE} due to our use of a different evaluation backbone~\cite{UniSpeechSpeakerVerification}.
\end{minipage}

\label{tab:dns1}
\end{table*}

\subsection{Ablation Study}
\subsubsection{Effects of Dry-Target Finetuning}
We investigate the effects of dry-target finetuning on the DNS1 with-reverb subset, with results reported in Table~\ref{tab:ablation}. For brevity, we only report UTMOS, SpkSim, and dWER here. To clarify, DeWavLM and DeWavLM-R are evaluated by synthesizing waveforms from phonetic representations using a phonetic vocoder trained on clean speech for reconstruction.

Comparisons between DeWavLM and DeWavLM-R, as well as between PASE and PASE-R, show that dry-target supervision consistently improves perceptual quality, evidenced by substantial gains in UTMOS (2.42 to 3.98 and 1.61 to 3.23). In addition, the further reduction in dWER (10.30\% to 8.69\% and 9.78\% to 8.01\%) indicates that PASE's semantic enhancement mechanism remains effective under severe reverberation, with semantic fidelity further improved using dry targets.

\begin{table}[t]
\centering
\caption{Comparison results on the simulated test set.}
\setlength{\tabcolsep}{0.6mm}
\resizebox{\linewidth}{!}{
\begin{tabular}{ccccccc}
\toprule
Model & DNSMOS $\uparrow$ & UTMOS $\uparrow$ & SBS $\uparrow$ & LPS $\uparrow$ & SpkSim $\uparrow$ & WER (\%) $\downarrow$ \\ \midrule
Noisy & 1.34 & 1.42 & 0.56 & 0.53 & 0.54 & 14.74 \\
Clean & 3.35 & 4.28 & 1.00 & 1.00 & 1.00 & 3.14 \\ \midrule
TF-GridNet & 2.87 & 2.22 & 0.77 & 0.82 & \textbf{0.73} & 13.46 \\
FlowSE & 3.28 & 3.50 & 0.76 & 0.74 & 0.59 & 27.84 \\
PASE & 2.96 & 2.44 & 0.82 & 0.86 & 0.53 & 13.95 \\
SenSE & \textbf{3.39} & \underline{3.88} & \underline{0.83} & \underline{0.88} & \underline{0.68} & \underline{12.73} \\
AES-V2 & 3.33 & 3.85 & 0.74 & 0.76 & 0.49 & 25.97 \\
StuPASE & \underline{3.37} & \textbf{4.08} & \textbf{0.85} & \textbf{0.90} & \underline{0.68} & \textbf{11.57} \\ \bottomrule
\end{tabular}
}
\label{tab:libri}
\end{table}

\subsubsection{Effects of Flow-Matching}
We evaluate the effectiveness of the flow-matching module and two key design choices: (i) using enhanced phonetic representations as semantic conditions, and (ii) the speech-infilling training strategy. Both designs target semantic guidance.

We first verify that our independently trained Mel vocoder can generate studio-quality speech. As shown in Table~\ref{tab:ablation}, it achieves near-perfect reconstruction (UTMOS 3.85, SpkSim 0.96, and dWER 0.92\%). 
Comparing PASE-R and StuPASE shows that the flow matching module substantially improves UTMOS (3.23 to 4.01) and further reduces dWER (8.01\% to 7.89\%), with minor SpkSim degradation (0.80 to 0.74), highlighting its advantage in producing high-fidelity speech.

Ablation studies further reveal the critical role of semantic guidance. Using noisy semantic conditions extracted from the original WavLM (\emph{w/ noisy semantic}) degrades all metrics, especially dWER (8.21\% to 19.79\%). Removing semantic conditions (\emph{w/o semantic}) further drops UTMOS (3.75 to 3.25) and increases dWER (19.79\% to 36.36\%). 
Omitting the masking mechanism in speech-infilling training (\emph{w/o masking}) mainly reduces UTMOS (4.01 to 3.82), with minimal changes in SpkSim and dWER.
These results confirm that semantic guidance is essential for both high perceptual quality and low hallucinations: the purified semantic condition from DeWavLM-R provides reliable linguistic information and a clean source for generation, while masking facilitates its full utilization.

\subsection{Comparison with Baselines}
\label{sec:comparison}
\subsubsection{Objective Results on the DNS1 Test Set}
Table~\ref{tab:dns1} summarizes the comparison results on the DNS1 test set under both no-reverb and with-reverb conditions. 
Under no-reverb, StuPASE achieves top-tier performance with a UTMOS of 3.99, outperforming most baselines including FlowSE and SenSE, and closely matching that of AES-V2. It also delivers superior linguistic integrity, with a low dWER of 2.84\%.

In the with-reverb scenario, StuPASE attains the highest UTMOS (4.01) along with strong SBS (0.86) and LPS (0.92), while maintaining the lowest dWER (7.89\%) among all models, indicating its low-hallucination properties. In contrast, FlowSE, SenSE, and AES-V2 exhibit substantially higher dWER under reverberation, reflecting more pronounced hallucinations despite relatively high perceptual scores. These results demonstrate that StuPASE can generate perceptually high-quality speech with accurate linguistic content even under challenging reverberation.
Compared to SenSE, StuPASE achieves higher perceptual quality and linguistic integrity with only slightly lower speaker similarity, yet benefits from a simpler and more efficient framework.

\subsubsection{Objective Results on the Simulated Test Set}
Table~\ref{tab:libri} presents the comparison results on the simulated test set, which contains a broader range of noise types and reverberation scenarios. Consistent with the findings on the DNS1 set, StuPASE demonstrates clear advantages over the baselines, achieving the highest UTMOS (4.08), SBS (0.85), LPS (0.90), and lowest WER (11.57\%), while maintaining competitive SpkSim (0.68). The strong performance across diverse acoustic conditions indicates that StuPASE exhibits robust generalization to more challenging, unseen environments.

\begin{figure}[t]
    \centering
    \includegraphics[width=\linewidth]{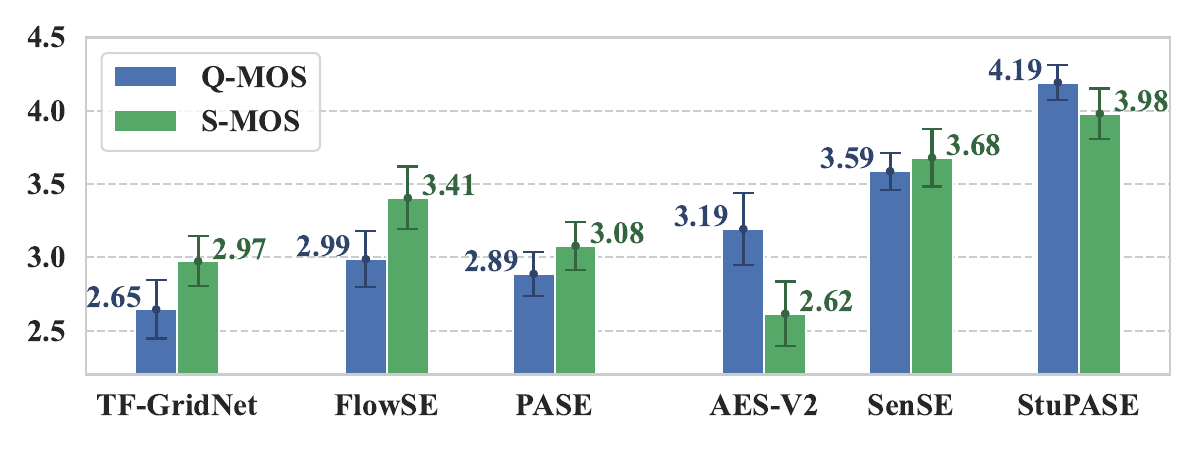}
    \caption{Subjective evaluation results for Q-MOS and S-MOS, with error bars indicating 95\% confidence intervals.}
    \label{fig:mos}
\end{figure}

\subsubsection{Subjective Results}
Fig.~\ref{fig:mos} presents the subjective evaluation results. 
Among all models, StuPASE achieves the best perceptual quality (Q-MOS 4.19) and speaker similarity (S-MOS 3.98), significantly outperforming the second-best SenSE (Q-MOS 3.59; S-MOS 3.68). Notably, despite identical objective SpkSim scores (0.68 in Table~\ref{tab:libri}), StuPASE achieves higher subjective S-MOS, indicating that its advantages are more pronounced to human listeners. These findings further validate the effectiveness of StuPASE in improving both perceptual quality and speaker consistency.

\section{Conclusion}
We propose StuPASE, a generative SE framework built on PASE. By using dry-target finetuning and replacing the GAN-based generative module with a flow-matching one, StuPASE achieves studio-level quality while maintaining low hallucination. Experiments show it consistently outperforms SOTA methods, advancing SE toward higher reliability and quality.

\section{Acknowledgments}
This work was supported by the National Natural Science Foundation of China (Grant No. 12274221) and the Yangtze River Delta Science and Technology Innovation Community Joint Research Project (Grant No. 2024CSJGG1100).

\section{Generative AI Use Disclosure}
The authors did not rely on generative AI to create the core content, analyses, or results of this paper. AI tools were only used sparingly for minor language refinement and editorial improvements. The authors take full responsibility for all scientific claims and content presented.

\bibliographystyle{IEEEtran}
\bibliography{ref}

\end{document}